\documentstyle[12pt]{article}
\catcode`\@=11

\@addtoreset{equation}{section}
\def\theequation{\arabic{section}.\arabic{equation}}

\catcode`\@=11

\def\section{\@startsection{section}{1}{\z@}{3.5ex plus 1ex minus
   .2ex}{2.3ex plus .2ex}{\large\bf}}

\def\eqnarray{\let\@currentlabel=\theequation\refstepcounter{equation}
    \global\@eqnswtrue
    \global\@eqcnt\z@\tabskip\@centering\let\\=\@eqncr
    $$\halign to \displaywidth\bgroup\@eqnsel\hskip\@centering
      $\displaystyle\tabskip\z@{##}$&\global\@eqcnt\@ne
       \hfil${{}##{}}$\hfil
      &\global\@eqcnt\tw@ $\displaystyle\tabskip\z@{##}$\hfil
       \tabskip\@centering&\llap{##}\tabskip\z@\cr}
\def\lefteqn#1{\hbox to 4\arraycolsep{$\displaystyle #1$\hss}}
\def\thesection{\arabic{section}.}

\def\appendix{\setcounter{section}{0}
        \def\thesection{Appendix.}
        \def\theequation{\Alph{section}.\arabic{equation}}}

\long\def\@makefntext#1{\parindent 0cm\noindent
\hbox to 1em{\hss$^{\@thefnmark}$}#1}
\def\IR{{\hbox{{\rm I}\kern-.2em\hbox{\rm R}}}}
\def\IH{{\hbox{{\rm I}\kern-.2em\hbox{\rm H}}}}
\def\IC{{\ \hbox{{\rm I}\kern-.6em\hbox{\bf C}}}}
\def\IZ{{\hbox{{\rm Z}\kern-.4em\hbox{\rm Z}}}}
\def\rref#1{(\ref{#1})}

\newcommand{\beq}{\begin{equation}}
\newcommand{\eeq}{\end{equation}}
\newcommand{\obs}{{\cal O}}

\newcommand{\limh}{\lim_{\hbar\rightarrow0}}

\newtheorem{Def}{Definition}

\newtheorem{result}[Def]{Result}
\newcommand{\PRD}[1]{{\sl Phys.~Rev.}~{\bf D#1}}

\newcommand{\JMP}[1]{{\sl J.~Math.~Phys.}~{\bf #1}}
\begin{document}
%
%
%
%
\def\citen#1{%
\edef\@tempa{\@ignspaftercomma,#1, \@end, }
\edef\@tempa{\expandafter\@ignendcommas\@tempa\@end}%
\if@filesw \immediate \write \@auxout {\string \citation {\@tempa}}\fi
\@tempcntb\m@ne \let\@h@ld\relax \let\@citea\@empty
\@for \@citeb:=\@tempa\do {\@cmpresscites}%
\@h@ld}
%
\def\@ignspaftercomma#1, {\ifx\@end#1\@empty\else
   #1,\expandafter\@ignspaftercomma\fi}
\def\@ignendcommas,#1,\@end{#1}
%
%
\def\@cmpresscites{%
 \expandafter\let \expandafter\@B@citeB \csname b@\@citeb \endcsname
 \ifx\@B@citeB\relax 
    \@h@ld\@citea\@tempcntb\m@ne{\bf ?}%
    \@warning {Citation `\@citeb ' on page \thepage \space undefined}%
 \else
    \@tempcnta\@tempcntb \advance\@tempcnta\@ne
    \setbox\z@\hbox\bgroup 
    \ifnum\z@<0\@B@citeB \relax
       \egroup \@tempcntb\@B@citeB \relax
       \else \egroup \@tempcntb\m@ne \fi
    \ifnum\@tempcnta=\@tempcntb 
       \ifx\@h@ld\relax 
          \edef \@h@ld{\@citea\@B@citeB}%
       \else 
          \edef\@h@ld{\hbox{--}\penalty\@highpenalty \@B@citeB}%
       \fi
    \else   
       \@h@ld \@citea \@B@citeB \let\@h@ld\relax
 \fi\fi%
 \let\@citea\@citepunct
}
%
\def\@citepunct{,\penalty\@highpenalty\hskip.13em plus.1em minus.1em}%
%
%
\def\@citex[#1]#2{\@cite{\citen{#2}}{#1}}%
%
%
\def\@cite#1#2{\leavevmode\unskip
  \ifnum\lastpenalty=\z@ \penalty\@highpenalty \fi 
  \ [{\multiply\@highpenalty 3 #1
      \if@tempswa,\penalty\@highpenalty\ #2\fi 
    }]\spacefactor\@m}
\let\nocitecount\relax  
%
\begin{titlepage}
\vspace{.5in}
\begin{flushright}
\end{flushright}
\vspace{.5in}
\begin{center}
{\Large\bf
 A Single Particle Interpretation\\[1ex]
of Relativistic Quantum Mechanics\\[1ex]
 in 1+1 Dimensions}\\
\vspace{.4in}
{R.~C{\sc osgrove}\\
       {\small\it Department of Physics}\\
       {\small\it University of California}\\
       {\small\it Davis, CA 95616}\\{\small\it USA}}\\
       {\small\it email: cosgrove@bethe.ucdavis.edu}
\end{center}

\vspace{.5in}
\begin{center}
\begin{minipage}{5in}
\begin{center}
{\large\bf Abstract}
\end{center}
{\small
The relativistic free particle system in 1+1 dimensions is formulated as a
``bi-Hamiltonian system''.  One Hamiltonian
generates ordinary time translations, and another generates (essentially)
boosts.  Any observer, accelerated or not, sees evolution as
the continuous unfolding of a canonical transformation which may be described
using the two Hamiltonians.  When the system is quantized both Hamiltonians
become Hermitian operators in the standard positive definite inner product.
Hence, each observer sees the evolution of the wave function as the
continuous unfolding of a unitary transformation in the standard positive
definite inner product.
The result appears to be a consistent single particle interpretation of
relativistic quantum mechanics.  This interpretation has the feature that
the wave function is observer dependent, and observables have non-local
character, similar to what one might expect in quantum gravity.
}
\end{minipage}
\end{center}
\end{titlepage}

Traditionally it has not been possible to extend the single particle
interpretation
of non-relativistic quantum mechanics to
relativistic quantum mechanics.  The main problem
has been the lack of a positive definite inner product associated with a
locally conserved current.
The necessity that the probability density transform
as the zeroth component of a 4-vector seems to preclude such inner product.

It is shown in this paper that
the Heisenberg picture formalism developed in
\cite{Cos} (motivated by Rovelli's ``evolving constants of the motion''
\cite{Rov1})
leads to a relativistic quantum mechanics in 1+1 dimensions which is a
natural extension of non-relativistic quantum mechanics, and which includes
a new notion of conserved current.  To make contact with the problems of
ordinary relativistic quantum mechanics it is necessary to switch to the
Schr\"{o}dinger picture.  It is found that doing this requires a paradigm
shift;  the wavefunction must be associated with an observer-system pair,
instead of with a system alone.

As has been suggested by Rovelli \cite{Rov2}, perhaps the notion of an
observer-independent
state of a system is flawed in a way analogous to the flaw in the notion of
observer-independent simultaneity.
I implement this concept in the simple example of the
quantum mechanics of the relativistic free particle by associating different
observers (accelerated or not) with different sequences of flat spacelike
submanifolds of Minkowski space, the submanifolds perpendicular to their
worldlines.
The notion of conserved current becomes the notion of consistent
unitary evolution for different observers.  This may be summarized by saying
that we regard the important object of the
formalism to be not a wave function
which is a mapping of spacetime to the complex numbers, but rather  a
mapping of the space of flat spacelike submanifolds of Minkowski space to
wavefunctions on space.

We will find that the physical consequence of adopting the above point
of view is that observables have non-local character; it is important for the
interpretation that detectors not be infinitesimal.  This makes the result
interesting from the point of view of quantum gravity, which is expected to
have this feature.  Whether or not this theory can be sensibly ``second
quantized'', and whether or not the results would agree with experiment is
not discussed.  It is not clear if the type of non-locality predicted would
be excluded by existing experimental data.  However, our results appear to be
equivalent to Fleming's  hyperplane dependent relativistic
quantum mechanics \cite{Fle89}, so that his discussion
\cite{Fle64,Fle,Fle89,Fle87} is relevant.


\section{Bi-Hamiltonian System}

In general relativity, asking what is the ``time'' should be replaced by
asking what is the ``spacelike submanifold''.  This presents a
problem in the interpretation of quantum mechanics; the standard
interpretations of quantum mechanics use the assumption that the ``time''
is a real number.
Because the technical problems of general relativity are so
immense,  it is useful to note that there is an analogous interpretation
of time in special relativity.  In special relativity asking what is the
``time'' should be replaced by asking what is the ``flat spacelike
submanifold in Minkowski space''.  These planes are parametrized by Lorentz
transformations and ordinary time translations.  If we restrict to the
(1+1)-dimensional case and fix a reference coordinate system, then the
straight spacelike lines (flat spacelike submanifolds)  can be parametrized by
the ordinary times where the lines cross the time axis, $t$, and a
boost parameter, $u$, which parametrizes the angles  that the lines
make with the horizontal.  Specifically let $\alpha$ be the angle that a line
makes with the horizontal and take
\beq
u=\frac{tan\ \alpha}{\sqrt{1-tan^{2}\ \alpha}}=
\frac{\beta}{\sqrt{1-\beta^{2}}},
\eeq
which ranges $(-\infty,\infty)$.

An observer determines a continuous ordered sequence of planes,
the planes perpendicular to the 4-velocity of its timelike path.
Hence, an observer defines a path in $\IR\times\IR$ by the evolution of the
parameters $t$ and $u$ associated with it.
We will refer to $\IR\times\IR$ as the space of global time.  Clearly the
concept of global time can be extended to higher dimensional special
relativity; the space of global time will be higher dimensional.  In
reference \cite{Cos} the concept of a space of global time
is discussed in a more general way, with reference to quantum
gravity.

Let $q(t,u)$ and $p(t,u)$ be the position and momentum in
the coordinate system boosted relative to the reference coordinate system
as determined by $u$, at the time determined by $t$.
Then we have
the following:
\begin{result}
\label{result1}
Any observer, that is any path in the space of global time:
\begin{eqnarray}
\phi : R&\rightarrow&\IR\times\IR = space\ of\ global\ time,\\
\tau&\mapsto&   (t(\tau),u(\tau)),
\end{eqnarray}
sees evolution of the (1+1)-dimensional free particle parameters
$q(t(\tau),u(\tau))$ and $p(t(\tau),u(\tau))$ as the
continuous unfolding of a canonical transformation.  There are two Hamiltonians
\begin{eqnarray}
H_{t}&=&\sqrt{1+u^{2}}\sqrt{m^{2}+p^{2}},\ \ and\nonumber\\
H_{u}&=&\frac{tu}{\sqrt{1+u^{2}}}\sqrt{m^{2}+p^{2}}+
\frac{1}{\sqrt{1+u^{2}}}q\sqrt{m^{2}+p^{2}}-tp
\end{eqnarray}
which together determine the evolution seen by any observer according to
\begin{eqnarray}
\label{bih}
dq&=&dt\{q,H_{t}\}_{pb}+
du\{q,H_{u}\}_{pb},\ \ \nonumber\\
dp&=&dt\{p,H_{t}\}_{pb}+
du\{p,H_{u}\}_{pb}.
\end{eqnarray}
\end{result}

The first Hamiltonian, $H_{t}$, describes the evolution as seen by an
observer who does
not accelerate (i.e. $t$ changes while $u$ remains fixed).  The second,
$H_{u}$, describes the evolution seen by a ``highly'' accelerated observer
(i.e. $u$ changes while $t$ remains fixed).
Hence, we will say that the (1+1)-dimensional relativistic free particle
system is a ``bi-Hamiltonian system''.

To demonstrate Result~\ref{result1}, work out the expressions for
$q(t,u)$ and $p(t,u)$ in terms of the ``initial data''
$q(0,0)=q_{0}$ and $p(0,0)=p_{0}$.  The calculation is not illuminating
and will be omitted.  The result is
\begin{eqnarray}
\label{qpf}
q(t,u)&=&q_{0}\frac{\sqrt{m^{2}+p_{0}^{2}}}{\sqrt{1+u^{2}}
\sqrt{m^{2}+p_{0}^{2}}-
up_{0}}+\frac{tp_{0}}{\sqrt{1+u^{2}}\sqrt{m^{2}+p_{0}^{2}}-
up_{0}}-tu,\ \ and\nonumber\\
p(t,u)&=&\sqrt{1+u^{2}}p_{0}-u\sqrt{m^{2}+p_{0}^{2}}.
\end{eqnarray}
It may be verified by direct substitution that the evolution determined
by \rref{qpf} satisfies equation \rref{bih}.
(Again the calculation is not illuminating and will
be omitted.)  Hence, the evolution is by canonical transformation and
Result~\ref{result1} is demonstrated.  Of course it may also be verified that
$\{q(t,u),p(t,u)\}_{pb}=1$.

\section{Quantization After Evolution}

For an ordinary Hamiltonian system, evolution is given by the Hamiltonian
equations of motion:
\beq
\frac{dq}{dt}=\{q,H\},\ \ \frac{dp}{dt}=\{p,H\}.
\eeq
If at some fixed initial time $t_{0}$ we have $q(t_{0})=q_{0}$ and
$p(t_{0})=p_{0}$, then at each later time, $t$, $q(t)$ and $p(t)$ can be
expressed as functions of $q_{0}$ and $p_{0}$:
\begin{eqnarray}
q(t)&=&Q(t,q_{0},p_{0}),\nonumber\\
p(t)&=&P(t,q_{0},p_{0}).
\end{eqnarray}
The evolution is by canonical transformation so that
\beq
\{q(t),p(t)\}=\{q_{0},p_{0}\} \stackrel{def}{=} 1.
\eeq

Ordinarily in quantizing a Hamiltonian system we make $q_{0}$, $p_{0}$ and
$H$ into
Hermitian operators $\hat{q}_{0}$, $\hat{p}_{0}$ and $\hat{H}$, and evolve
by means of the Heisenberg equations of motion:
\beq
\frac{d\hat{q}}{dt}=\frac{1}{i\hbar}[\hat{q},\hat{H}],\ \
\frac{d\hat{p}}{dt}=\frac{1}{i\hbar}[\hat{p},\hat{H}].
\eeq
If at some fixed time $t_{0}$ we have $\hat{q}(t_{0})=\hat{q}_{0}$
and $\hat{p}(t_{0})=\hat{p}_{0}$, then at each later time, $t$, $\hat{q}(t)$
and $\hat{p}(t)$ can be expressed as functions of $\hat{q}_{0}$ and
$\hat{p}_{0}$:
\begin{eqnarray}
\hat{q}(t)=\hat{Q}(t,\hat{q}_{0},\hat{p}_{0}),\nonumber\\
\hat{p}(t)=\hat{P}(t,\hat{q}_{0},\hat{p}_{0}).
\end{eqnarray}
The evolution is by unitary transformation so that
\beq
\frac{1}{i\hbar}[\hat{q}(t),\hat{p}(t)]=\frac{1}{i\hbar}[
\hat{q}_{0},\hat{p}_{0}]\stackrel{def}{=} 1.
\eeq

We will refer to the function $\obs$ on phase space (for example $q_{0}$)
which
corresponds to the operator $\hat{\obs}$ (for example $\hat{q}_{0})$ as the
classical limit of
$\hat{\obs}$, or $\obs=\limh \hat{\obs}$.  In Heisenberg picture
quantization we
define the classical limit of
$\hat{q}_{0}$ to be $q_{0}$ and the classical limit of $\hat{p_{0}}$ to be
$p_{0}$.  That is, the quantization is
carried out at some fixed time $t_{0}$.  The use of the Heisenberg equations
of motion to find $\hat{q}(t)$ and $\hat{p}(t)$ for $t>t_{0}$ is supposed to
ensure that
$\limh \hat{q}(t) = q(t)$
and $\limh \hat{p}(t) =p(t)$.  This means that the functions
$\hat{Q}(t,\ast,\ast)$ and $\hat{P}(t,\ast,\ast)$  of $\hat{q}_{0}$
and
$\hat{p}_{0}$ are obtained from the functions $Q(t,\ast,\ast)$ and
$P(t,\ast,\ast)$  of $q_{0}$ and $p_{0}$
by substituting $\hat{q}_{0}$  and $\hat{p}_{0}$ for $q_{0}$ and $p_{0}$
and choosing ``the correct operator ordering''. (Of course the choice
of operator ordering does not effect the classical limit.)  Hence, one
might attempt to quantize the system directly by making this substitution,
without reference to the Heisenberg equations of motion.
This operation will be called ``quantization after evolution'', because
the evolution
problem is solved classically and the whole ``already evolved classical
system'' is quantized at once. It is closely related to Rovelli's ``evolving
constants'' picture of dynamics \cite{Rov1}.  Examples of this construction
have been worked out by Carlip \cite{Car} in the context of
comparing different quantizations.

The obvious approach to quantizing our bi-Hamiltonian system
is to turn $q_{0}$, $p_{0}$, $H_{t}(t,u)$ and
$H_{u}(t, u)$ into hermitian operators $\hat{q}_{0}$, $\hat{p}_{0}$,
$\hat{H}_{t}(t,u)$ and $\hat{H}_{u}(t,u)$, and evolve
to arbitrary global
time $(t, u)$ using the pair of Heisenberg equations of
motion.  There is, however, a potential problem here.  Due to operator
ordering ambiguities, evolving from the global time $(t_{0}, u_{0})$ to
the global
time $(t, u)$ by different paths in the space of global time may
yield
different results.  That is to say, an observer who moves from global
time $(t_{0},u_{0})$ to global time $(t,u)$ by
accelerating from speed $v_{0}$ and then decelerating back to speed
$v_{0}$ may find
different operators $\hat{q}(t,u)$ and $\hat{p}(t,u)$ then
an observer who makes the trip
(through the space of global time) by staying at the constant velocity
$v_{0}$.
If the name ``space of global time'' is worth its salt, then all observers
should agree on $\hat{q}(t,u)$ and $\hat{p}(t,u)$.

Note now that the operation of quantization after evolution, applied to the
bi-Hamiltonian system, is inherently path independent.  If we fix
some ``initial'' global time $(t_{0},u_{0})$, then for each
$(t,u)$ we find that $q(t,u)$ and $p(t,u)$ are
expressed as functions of $q(t_{0},u_{0})=q_{0}$ and
$p(t_{0},u_{0})=p_{0}$:
\begin{eqnarray}
\label{ecm}
q(t,u)&=&Q(t,u,q_{0},p_{0})\nonumber\\
p(t,u)&=&P(t,u,q_{0},p_{0}).
\end{eqnarray}
The functions $Q$ and $P$ are given explicitly in equation \rref{qpf}.
Quantization after evolution involves substituting $\hat{q}_{0}$ and
$\hat{p}_{0}$  for $q_{0}$ and $p_{0}$ in the functions $Q$
and $P$ and
choosing the operator orderings so that $\hat{q}(t,u)$ and
$\hat{p}(t,u)$ are hermitian and related
to $\hat{q}_{0}$ and $\hat{p}_{0}$ by a  unitary transformation for all
$(t,u)$.  The important point,
however, is that no matter what operator orderings are chosen, the above
substitution yields  unique values for $\hat{q}(t,u)$ and
$\hat{p}(t,u)$ for each $(t,u)$, and
therefore avoids the issue of path dependence mentioned above.

Now let us quantize the bi-Hamiltonian system using quantization after
evolution.
We substitute the operators $\hat{q}_{0}$ and
$\hat{p}_{0}$ for $q_{0}$ and $p_{0}$ in the functions $Q$ and $P$
(equation \rref{qpf}), and choose the most obvious operator orderings which
make $\hat{q}(t,u)$ and $\hat{p}(t,u)$ Hermitian in the
inner product
\beq
\int \psi^{\ast}(q,t,u)\psi(q,t,u) dq
\eeq
(where $\hat{q}_{0}$ is represented by multiplication by $q_{0}$ and
$\hat{p}_{0}$ is
represented by  $\frac{\hbar}{i}\frac{\partial}{\partial q_{0}}$).
The result is
\begin{eqnarray}
\label{qpo}
\hat{q}(t,u)&=&\frac{1}{2}\hat{q}_{0}\frac{\sqrt{m^{2}+
\hat{p}_{0}^{2}}}
{\sqrt{1+u^{2}}\sqrt{m^{2}+p_{0}^{2}}-
u\hat{p}_{0}}\nonumber\\
&+&\frac{1}{2}\frac{\sqrt{m^{2}+\hat{p}_{0}^{2}}}
{\sqrt{1+u^{2}}\sqrt{m^{2}+p_{0}^{2}}-
u\hat{p}_{0}}\hat{q}_{0}+
\frac{t\hat{p}_{0}}{\sqrt{1+u^{2}}
\sqrt{m^{2}+\hat{p}_{0}^{2}}-
u\hat{p}_{0}}-tu,\ \ and\nonumber\\
\hat{p}(t,u)&=&\sqrt{1+u^{2}}\hat{p}_{0}-u
\sqrt{m^{2}+\hat{p}_{0}^{2}}.
\end{eqnarray}
The operator $\sqrt{m^{2}+\hat{p}_{0}^{2}}$ may be defined by means of
spectral decomposition.  To get a ``single particle interpretation'' only the
positive (or negative) root of the eigenvalues of the operator
$m^{2}+\hat{p}_{0}^{2}$ should be retained. This is in some sense justified
by the fact that the classical Hamiltonian is the positive branch of the
square root of $m^{2}+p_{0}^{2}$. The definition of the square root operator
in this way has been discussed in \cite{Lam}.

We do not yet know if $\hat{q}(t,u)$ and $\hat{p}(t,u)$
are related to $\hat{q}_{0}$ and $\hat{p}_{0}$ by a unitary
transformation for all $(t,u)$.  To verify this, we make
$H_{t}$ and $H_{u}$ into
Hermitian operators $\hat{H}_{t}$ and $\hat{H}_{u}$ by choosing the most
obvious operator orderings which make them Hermitian:
\begin{eqnarray}
\label{qhams}
\hat{H}_{t}&=&\sqrt{1+u^{2}}\sqrt{m^{2}+\hat{p}^{2}},\ \ and\nonumber\\
\hat{H}_{u}&=&\frac{tu}{\sqrt{1+u^{2}}}\sqrt{m^{2}+\hat{p}^{2}}+
\frac{1}{2}\frac{1}{\sqrt{1+u^{2}}}\hat{q}\sqrt{m^{2}+\hat{p}^{2}}+
\frac{1}{2}\frac{1}{\sqrt{1+u^{2}}}\sqrt{m^{2}+\hat{p}^{2}}\hat{q}
-t\hat{p}.
\end{eqnarray}
It is straightforward to verify that the expressions for $\hat{q}(t,u)$ and
$\hat{p}(t,u)$ \rref{qpo} satisfy the
Heisenberg equations of motion
\begin{eqnarray}
\frac{\partial\hat{q}}{\partial t}=\frac{1}{i\hbar}[\hat{q},\hat{H}_{t}]
&,&\ \ \frac{\partial\hat{p}}{\partial t}=\frac{1}{i\hbar}[\hat{p},
\hat{H}_{t}],\nonumber\\
\frac{\partial\hat{q}}{\partial u}=\frac{1}{i\hbar}[\hat{q},\hat{H}_{u}]
&,&\ \ \frac{\partial\hat{p}}{\partial u}=\frac{1}{i\hbar}[\hat{p},
\hat{H}_{u}],
\end{eqnarray}
so that operator evolution is by unitary transformation.

We have now demonstrated the following:
\begin{result}
Any observer, that is any path in the space of global time:
\begin{eqnarray}
\phi : R&\rightarrow&\IR\times\IR = space\ of\ global\ time,\\
\tau&\mapsto&   (t(\tau),u(\tau)),
\end{eqnarray}
sees evolution of the free particle operators $\hat{q}(t(\tau), u(\tau))$
and $\hat{p}(t(\tau), u(\tau))$
as the continuous unfolding of a unitary transformation in the standard
positive definite inner product,
\beq
\int \psi^{\ast}(q,t,u) \psi(q,t,u) dq
\eeq
(where $\hat{q}_{0}$ is represented by multiplication with $q_{0}$ and
$\hat{p}_{0}$ is represented by $\frac{\hbar}{i}\frac{\partial}{
\partial q_{0}}$).
There are two Hamiltonians \rref{qhams} which together
determine this evolution for any observer according to
\begin{eqnarray}
\label{biho}
d\hat{q}&=&dt\frac{1}{i\hbar}[\hat{q},\hat{H}_{t}]+
du\frac{1}{i\hbar}[\hat{q},\hat{H}_{u}],\ \ \nonumber\\
d\hat{p}&=&dt\frac{1}{i\hbar}[\hat{p},\hat{H}_{t}]+
du\frac{1}{i\hbar}[\hat{p},\hat{H}_{u}].
\end{eqnarray}
The evolution is consistent in the sense that all observers agree on the
operators $\hat{q}(t,u)$ and $\hat{p}(t,u)$ for each
global time $(t,u)$.
\end{result}

The first Hamiltonian, $\hat{H}_{t}$, describes the
operator
evolution as seen by an observer who does not accelerate (i.e. $t$ changes
while $u$ remains fixed).  The second, $\hat{H}_{u}$, describes the
evolution seen by
a ``highly'' accelerated observer (i.e. $u$ changes while $t$ remains fixed).
For an unaccelerated observer there is a wave function on spacetime which
satisfies the square root Sch\"{o}dinger equation.  This equation has been
discussed in \cite{Lam} and in \cite{Smi}.  In \cite{Lam} the positive root
is taken so
that there are no negative frequency modes.  Strictly speaking, it is only
in this way that a ``single particle interpretation'' is obtained, since the
negative frequency modes are generally considered to represent different
particles.  This gives rise to many interesting issues concerning the
localization of particles and position operators \cite{Pry,Mol,Jor,New,Fle64}.
As mentioned below, these issues lead to an alternative motivation for the
present work.
It is also possible to relax the restriction to a single particle and allow
negative energy modes; the eigenvalues of the square root operator acting
on the negative energy modes are simply taken to be negative.  However,
as discussed in \cite{Smi}, there is some question as to the correct way to
introduce interactions between the positive and negative energy modes in
the setting of the square root Schr\"{o}dinger equation.

\section{Interpretation of Results}

We have defined operators $\hat{q}(t,u)$ and $\hat{p}(t,u)$
on the space of global time, $\IR\times\IR$, which ``evolve'' by unitary
transformation along any path in the space of global time, with the standard
positive definite inner product.  (The reader is reminded that the space of
global time in the example of this paper is the space of flat spacelike
submanifolds of Minkowski space.)
The relevant path in the space of global time is singled out by the choice
of observer.  It is the set of submanifolds perpendicular to the observer's
worldline.  This allows us to make physical predictions
of the following form: an observer may measure an observable at global
time $(t_{0},u_{0})$, for example $\hat{q}(t_{0},u_{0})$, thus producing a
state $|\psi_{0}\rangle$ which is an eigenstate of
$\hat{q}(t_{0},u_{0})$,
\beq
\hat{q}(t_{0},u_{0})|\psi_{0}\rangle=q(t_{0},u_{0})|\psi_{0}\rangle;
\eeq
the observer may then evolve along any path in the space of global time
to $(t,u)$ (if $u\neq u_{0}$ this will involve acceleration), and predict the
outcome of its measurement of, for example, $\hat{q}(t,u)$.  The expectation
value for the observer's second measurement is
$\langle\psi_{0}|\hat{q}(t,u)|\psi_{0}\rangle$.

The above implies that different global times are regarded as different
``times''.
To understand the physical implication of this,
consider the case in which the
observer is located on the time axis of the reference coordinate system
when it makes its second measurement.  Then the different global times
$(t,u)$ and $(t,u^{\prime})$ find the observer at the same point
in spacetime.  Hence the fact that
$\hat{q}(t,u)\neq\hat{q}(t,u^{\prime})$ when $u\neq u^{\prime}$
implies that the
probability amplitude for the outcome of a position measurement depends
on the observer's velocity, not just on the observer's position in spacetime.
This reflects the fact that different velocities correspond to
different definitions of ``simultaneous'', that is to different spatial slices
through the observer's position in spacetime.  It is physically sensible as
long as the observer's detector is of finite size (not infinitesimal),
because a detector of finite size at global time
$(t,u)$ occupies a different region of spacetime
(different spatial slice)
than when it is at global time $(t,u^{\prime})$.  It follows that
measurements of the position
operators $\hat{q}(t,u)$ and $\hat{q}(t,u^{\prime})$ are genuinely
different physical measurements.
Although it would seem to be
impossible to determine a particle's position with an infinitesimal detector,
we should note that these statements break down for an infinitesimal detector;
the position in
spacetime of an infinitesimal detector located on the time axis is
independent of $u$.
In this sense, non-locality of physical observables plays an
important part in this formulation of relativistic quantum mechanics.
It is not immediately clear if this kind of non-locality is incompatible
with experimental results.  A careful study of the measurement
process is necessary.

To understand how this interpretation relates to the standard problems of
relativistic quantum mechanics it is necessary to switch to the
Schr\"{o}dinger picture.  Corresponding to each global time is an
instantaneous wave function defined on the spacelike submanifold associated
with the global time.  However, spacelike submanifolds which correspond
to different global times may intersect in spacetime.  There appears to be
no reason to believe
that the instantaneous wave functions defined on the submanifolds
will agree at the intersection points.  In the Appendix we show that in
general they do not.
Hence, it is not
possible to define the wave function as a mapping of spacetime to the
complex numbers.
Instead, the important object of
the formalism is a mapping of the space of global time (flat
spacelike planes in Minkowski space) to wave functions on space:
\beq
\Phi:\{\hbox{embeddings}\ \Sigma\hookrightarrow M\ |\ \hbox{spacelike\
and\ flat}\}\rightarrow\Sigma^{\ast},
\eeq
where $\Sigma^{\ast}$  is the complex dual of $\Sigma$, and in n+1
dimensions  $\Sigma$ is $\IR^{n}$.  An observer is then associated with a
sequence of instantaneous wave functions, different sequences for different
observers.  Only if the observer is
unaccelerated is it clear that this gives rise to a wave function on
spacetime.  (As mentioned earlier, this wave function will satisfy the
square root Schr\"{o}dinger equation.)

In essence, the observer has been ``relativized'' (as Smolin puts it
\cite{Smo}).
The state is not considered as an object associated with the
physical
system, but rather as an object associated with an observer observing a
physical system.\footnote{Stated another way, it is an object associated
with a set
of measurements that have been made together with a set of measurements
that can be made.  This seems to tie in with consistent histories
quantum mechanics.}  In
Rovelli's 1993 preprint ``On Quantum Mechanics'' \cite{Rov2}, he
suggests that the notion of  observer-independent state of a system
may be flawed in a way analogous to the flaw in the notion of
observer-independent simultaneity.
In the construction of the present paper, if $q(t,u)$ and
$q(t,u^{\prime})$ have different values when
$u\neq  u^{\prime}$,  then  the operators associated
with them, $\hat{q}(t,u)$ and $\hat{q}(t,u^{\prime})$,
are different.  This seems to be sensible and indeed necessary if the theory
is to have the correct classical limit.  Yet it means that, in the Heisenberg
picture, observers at the same point in spacetime moving at different
velocities naturally measure different operator observables.  Switching
to the Schr\"{o}dinger picture, it means that these observers are naturally
associated with different wavefunctions.  This is a consequence of the fact
that their notions of simultaneity are different.  Hence we are led to
accept the suggestion of Rovelli, and closely related suggestions of
Smolin \cite{Smo}, and Crane \cite{Cra}, and to propose this
somewhat unconventional relativistic quantum mechanics.\footnote{Certainly,
the result obtained here does not capture the full spirit of the work of
Rovelli, Smolin, and Crane.  The free particle system is not
sufficiently complex; it is not possible to consider a variety of
observer-system splits.  We have merely considered a variety of abstract
observers observing the free particle system.}

To evaluate the self-consistency of a theory of this form we must discuss
the issue of multiple observers.  Consider the free particle system with
two observers, Observer Number One and Observer Number Two.  In its
calculations, Observer Number One can always, if necessary, consider the
larger system of Observer Number Two interacting with the free particle.
As long as this is a sensible physical system, then Observer Number One will
not find any contradictions.  Hence, Observer Number One will not find a
contradiction as long as it doesn't ignore part of the physical system,
namely Observer Number Two, in its calculations.  Of course if Observer
Number Two never interacts with Observer Number One, and never interacts
with the free particle (except, possibly, to duplicate some of the measurements
made by Observer Number One),
then Observer Number One need only consider the
free particle system in its calculations.
Rovelli
has discussed interpretational issues of this type in \cite{Rov2}.  The
important point is that we avoid internal inconsistencies by studying
observer-system pairs, instead of systems in the abstract.  Another way to
say this is that we study sequences of measurements.
Hence, this approach is related to consistent histories quantum
mechanics.  (The relativistic quantum mechanics of the free particle
appears to be sensible in the consistent histories formalism \cite{Har,Whe}.)

The key element that allows this single particle interpretation to succeed
is the replacement of the notion of locally conserved current with the notion
of consistent unitary evolution for different observers.  For a classical
particle flux the answer to the question ``What is the probability for
finding a particle in a certain region of spacetime?'' is observer
independent.  This is the essence of the concept of locally conserved current.
However, an actual measurement determining if a particle is in a region of
spacetime consists of performing a measurement to determine if a particle is
in a region of space, and then continuing this measurement for some period
of time.  If different spatial slices are used to define simultaneity,
then a different experimental procedure is required.  Although classical
special relativity predicts that these two experimental procedures will
yield the same result, it is not clear that this equivalence should be
regarded as an essential element of a relativistic theory.  The version
of quantum mechanics presented here is proclaimed to be relativistic
because the quantization map
$(q(t,u)\mapsto\hat{q}(t,u),p(t,u)\mapsto\hat{p}(t,u))$ induces an isomorphism
between two representations of the group generated by Lorentz transformations
and time translations, one with representation space
$\{(q(t,u),p(t,u))\ |\ t,u\in\IR\}$ and the other with representation space
$\{(\hat{q}(t,u),\hat{p}(t,u))\ |\ t,u\in\IR\}$.  It is a matter of taste
whether or not this is enough to call the theory relativistic.
The present theory does not
predict that the two experimental procedures described above will yield
the same result.  In this version of relativistic quantum mechanics the
answer to the question ``What is the probability for finding a particle in
a certain region of spacetime?'' is observer dependent.  The closest
concept to that of locally conserved current that is allowed is that of
consistent unitary evolution for different observers:  stated in the
Heisenberg picture, all observers see unitary evolution of the operator
observables, and all observers present at a particular global time agree
on the operator observables;  stated in the Schr\"{o}dinger picture, all
observers see unitary evolution of their wavefunctions, and all observers
present at a particular global time have the same wavefunction.

In the late stages of this work I was introduced to the work of
Fleming.\footnote{I have Steve Carlip to thank for this.}  It appears that
the theory developed here is, in essence, an evolving constants
approach to Fleming's \cite{Fle89} hyperplane (flat spacelike submanifold)
dependent relativistic quantum mechanics.  In a series of papers
\cite{Fle64,Fle,Fle87,Fle89} Fleming has developed the theory of hyperplane
dependent operator observables.  He has even begun study of hyperplane
dependent quantum field theory \cite{Fle87}.  His work is motivated
by the study of position operators and the fact that the Newton-Wigner
position operator for a single particle (constructed out of only positive
frequency modes) is not relativistically invariant; if a particle is localized
on one hyperplane, then it is not localized on another that intersects the
first at the location of the particle.  In the present work, the emphasis
is on quantization after evolution as a way to understand quantization of
systems with space of global time not equal to $\IR$.  We hope that our
approach will help to clarify the interpretation of Fleming's work.
However, our main purpose is to build up ideas that may be applicable to
interpreting quantum gravity.

Smolin's discussion of quantum cosmology in \cite{Smo}, and Crane's
discussion of quantum gravity in \cite{Cra}, have much in
common with the above discussion.
The relativistic quantum mechanics presented here may be an example of the
type of situation we are faced with in
quantum gravity.  There may be a space of global time which
is something like the space of spacelike submanifolds.
(However, this particular space of global time
doesn't quite make sense in the context of quantum gravity.
For a more technical discussion
of the possibility of viewing gravity in this way see \cite{Cos,Kuk,Ish}.)
An observer may define a path in
this (infinite dimensional) space of global time.  If we could find a
natural way to relate an experimental apparatus to such a path, then the
understanding described here may provide a conceptually acceptable view of
quantum gravity, without structural modification of general relativity or
quantum mechanics.

\section{Appendix: Demonstration of Non-Locality}

To see that it is not possible to define the wave function as a mapping of
spacetime to the complex numbers, we will show that the wave functions on
two different spacelike submanifolds (i.e. at two different global times)
through the origin of the reference coordinate system
will not always agree at the origin.  The restriction to positive energy
states will be assumed.
Let $|p(t,u)\rangle$ be the basis eigenket
for the operator $\hat{p}(t,u)$.  From \rref{qpo} we find
\beq
\label{eig}
\hat{p}(t,u)|p(0,0)\rangle = (\sqrt{1+u^{2}}p-u\sqrt{m^{2}+p^{2}})|p(0,0)
\rangle.
\eeq
{}From \rref{eig} it follows that the basis eigenkets $|p(0,0)\rangle$ and
$|(\sqrt{1+u^{2}}p-u\sqrt{m^{2}+p^{2}})(t,u)\rangle$ agree up to phase:
\beq
\label{eig2}
|p(0,0)\rangle = |(\sqrt{1+u^{2}}p-u\sqrt{m^{2}+p^{2}})(t,u)\rangle e^
{if(p,t,u)}.
\eeq
{}From \rref{eig2} we can relate the momentum space wave functions
$\psi_{p}(p,0,0)$ and $\psi_{p}(p,t,u)$ as follows:
\begin{eqnarray}
\label{pwave}
\psi_{p}(p,0,0)&=&\langle p(0,0)|\psi\rangle\nonumber\\
&=&\langle(\sqrt{1+u^{2}}p-u\sqrt{m^{2}+p^{2}})(t,u)|\psi\rangle e^
{if(p,t,u)}\nonumber\\
&=&\psi_{p}(\sqrt{1+u^{2}}p-u\sqrt{m^{2}+p^{2}},t,u)e^{if(p,t,u)}.
\end{eqnarray}
Finally, we write the position space wave functions $\psi_{q}(q,0,0)$ and
$\psi_{q}(q^{\prime},t,u)$ in terms of the momentum space wave functions
and use \rref{pwave} to obtain:
\begin{eqnarray}
\label{qwave}
\psi_{q}(q,0,0)&=&\langle q(0,0)|\psi\rangle\nonumber\\
&=&\int dp\langle q(0,0)|p(0,0)\rangle\langle p(0,0)|\psi\rangle\nonumber\\
&=&\int dp\frac{1}{\sqrt{2\pi}}e^{ipq}\psi_{p}(p,0,0)\nonumber\\
&=&\int dp\frac{1}{\sqrt{2\pi}}e^{ipq}\psi_{p}(\sqrt{1+u^{2}}p-u\sqrt
{m^{2}+p^{2}},t,u)e^{if(p,t,u)}\ \ and,\nonumber\\
\psi_{q}(q^{\prime},t,u)&=&\langle q^{\prime}(t,u)|\psi\rangle\nonumber\\
&=&\int dp \langle q^{\prime}(t,u)|p(t,u)\rangle\langle p(t,u)|\psi\rangle
\nonumber\\
&=&\int dp\frac{1}{\sqrt{2\pi}}e^{ipq^{\prime}}\psi_{p}(p,t,u).
\end{eqnarray}
Note that while we could certainly introduce invariant measures in \rref{qwave}
without changing the below conclusion, this would not be
in the spirit of the interpretation of relativity adopted in the present work.
The wave function $\psi_{p}(p,t,u)$ is associated with a particular global
time,
and a particular global time is associated with a particular boost parameter.
Changing global times is viewed as evolution.  (This is the ``clean separation
of the dynamical evolution problem from the kinematical transformation
problem''  that Fleming speaks of. \cite{Fle89} )

The wave function at the origin of the reference coordinate system on the
spacelike submanifold $(t=0,u=0)$ is $\psi_{q}(0,0,0)$, for which from
\rref{qwave}
we have
\beq
\label{q0wave}
\psi_{q}(0,0,0)=\int \frac{dp}{\sqrt{2\pi}}\psi_{p}(\sqrt{1+u^{2}}p-u\sqrt
{m^{2}+p^{2}},0,u)e^{if(p,0,u)}.
\eeq
(Here we have used the fact that the left hand side is
independent of $t$ to choose $t=0$ on the right hand side.)
The wave function at the origin of the reference coordinate system on the
spacelike submanifold $(t=0,u)$ is $\psi_{q}(0,0,u)$, for which from
\rref{qwave}
we have
\beq
\label{quwave}
\psi_{q}(0,0,u)=\int \frac{dp}{\sqrt{2\pi}}\psi_{p}(p,0,u).
\eeq
This demonstrates that our formalism allows $\psi_{q}(0,0,0)$ and
$\psi_{q}(0,0,u)$ to be unequal.  For example, consider the special case
$\psi_{p}(0,0,u)=\delta(p)$.  Direct calculation gives
\begin{eqnarray}
\label{compare}
\psi_{q}(0,0,0)&=&\frac{1}{\sqrt{2\pi(1+u^{2})}}e^{if(um,0,u)},\ \ and
\nonumber\\
\psi_{q}(0,0,u)&=&\frac{1}{\sqrt{2\pi}}.
\end{eqnarray}
Thus it is demonstrated that the formalism developed here is non-local in the
sense that it possesses what Fleming calls ``hyperplane dependence''.

\begin{flushleft}
{\large\bf Acknowledgments}
\end{flushleft}

I would like to thank Steven Carlip for suggesting that I find an
example of a ``system with global time $\Gamma$'' (defined in \cite{Cos})
other than (possibly) gravity, and for other stimulating conversations.




\begin{thebibliography}{99}
\bibitem{Cos} R.\ Cosgrove, University of California, Davis preprint:
UCD-95-42, e-Print Archive: gr-qc/9511059
\bibitem{Rov1} C.\ Rovelli, \PRD{42} 2638 (1990); \PRD{43} 442 (1991).
\bibitem{Rov2} C.\ Rovelli, Pittsburgh University preprint: PITT-REL-02-94,
e-Print Archive: hep-th/9403015.
\bibitem{Fle89} G. N.\ Fleming and H.\ Bennett, Found. of Phys. {\bf 19}, 3
(1989).
\bibitem{Fle64} G. N.\ Fleming, Phys. Rev. {\bf 137}, 1B, 188 (1965).
\bibitem{Fle} G. N.\ Fleming, \JMP{7} 1959 (1966), \PRD{2} 542 (1970).
\bibitem{Fle87} G. N.\ Fleming, ``Hyperplane-dependent quantized Fields and
Lorentz Invariance,'' in ``Philosophical Foundations of Quantum Field Theory,''
edited by H. R.\ Brown and R.\ Harr\'e (Clarendon Press, Oxford, 1990).
\bibitem{Car} S.\ Carlip, \PRD{42} 2647 (1990),
S.\ Carlip and J.E.\ Nelson, \PRD{51} 5643 (1995).
\bibitem{Lam} C.\ L\"{a}mmerzahl, \JMP{34}~(9) 3918 (1993).
\bibitem{Smi} J.\ Smith, University of California, Davis preprint:
UCD-IIRPA-94-14.
\bibitem{Pry} M. H. L.\ Pryce, Proc. Roy. Soc. (London) {\bf A195}, 62 (1948).
\bibitem{Mol} C.\ M\o ller, Communication from Dublin Institute for Advanced
Studies {\bf A} No.5, 1949 (unpublished).
\bibitem{Jor} T. F.\ Jordon and N.\ Mukunda, Phys. Rev. {\bf 132}, 1842
(1963).
\bibitem{New} T. D.\ Newton and E. P.\ Wigner, {\sl Rev.~Mod.~Phys.}~{\bf 21}~
(3) 400 (1949).
\bibitem{Smo} L.\ Smolin, Pennsylvania State University preprint:
CGPG-95-8-7, e-Print Archive: gr-qc/9508064.
\bibitem{Cra} L.\ Crane, \JMP{36} (11) (1995).
\bibitem{Har} J. B.\ Hartle, University of California, Santa Barbara preprint:
UCSBTH92-21, lectures given at the 1992 Les Houches \'{E}cole d'\'{e}t\'{e},
Gravitation et Quantifications, July 9-17, 1992.
\bibitem{Whe} J.\ Whelan, \PRD{50} 6344 (1994).
\bibitem{Kuk} K.\ Kucha\v{r}, ``Time and Interpretations of Quantum Gravity,''
in 4th Canadian Conference on General Relativity and Relativistic Astrophysics
(World Scientific, Singapore, 1992)
\bibitem{Ish} C. J.\ Isham, ``Canonical Quantum Gravity and the Problem of
Time'' in ``Integrable Systems, Quantum Groups, and Quantum Field Theories'',
pp157--288, edited by L. A.\ Ibort and M. A.\ Rodriguez (Kluwer Academic
Publishers, London, 1993).
\end{thebibliography}
\end{document}